\documentclass[11pt,english,british,journal=langd5]{achemso}
\usepackage[latin9]{inputenc}
\usepackage{geometry}
\usepackage{array}
\usepackage{textcomp}
\usepackage{amsmath}
\usepackage{amssymb}
\usepackage{graphicx}
\usepackage{color}
\usepackage{graphics}
\usepackage{dcolumn}
\usepackage{upgreek}

\title{Non-Equilibrium Gibbs' Criterion for Completely Wetting Volatile Liquids}

\usepackage{hyperref}
\pdfoutput=1
\makeatother

\usepackage{babel}

\author{Yannis Tsoumpas}\email{itsoumpa@ulb.ac.be}
\author{Sam Dehaeck}
\affiliation[Universit\'e Libre de Bruxelles (ULB)]{Universit\'e Libre de Bruxelles, TIPs (Transfers, Interfaces and Processes), CP 165/67, Av. F.D. Roosevelt 50, 1050 Brussels, Belgium}
\author{Mariano Galvagno}
\affiliation[Loughborough University]{Loughborough University, Department of Mathematical Sciences, LE11 3TU Loughborough, U.K}

\author{Alexey Rednikov}
\affiliation[Universit\'e Libre de Bruxelles (ULB)]{Universit\'e Libre de Bruxelles, TIPs (Transfers, Interfaces and Processes), CP 165/67, Av. F.D. Roosevelt 50, 1050 Brussels, Belgium}
\author{Heidi Ottevaere}
\affiliation[Vrije Universiteit Brussel (VUB)]{Vrije Universiteit Brussel, Department of Applied Physics and Photonics (TONA-IR),\\ Pleinlaan 2, B1050 Brussels, Belgium}
\author{Uwe Thiele}
\affiliation[M\"unster University]{Center of Nonlinear Science (CeNoS), Westf\"alische Wilhelms Universit\"at M\"unster, Corrensstr. 2, D48149 M\"unster, Germany}
\affiliation[M\"unster University]{Institut f\"ur Theoretische Physik, Westf\"alische Wilhelms Universit\"at M\"unster, Wilhelm Klemm Str. 9, D48149 M\"unster, Germany}
\author{Pierre Colinet}\email{pcolinet@ulb.ac.be}
\affiliation[Universit\'e Libre de Bruxelles (ULB)]{Universit\'e Libre de Bruxelles, TIPs (Transfers, Interfaces and Processes), CP 165/67, Av. F.D. Roosevelt 50, 1050 Brussels, Belgium}

\makeatother

\usepackage{babel}
\begin{document}
\begin{abstract}
During the spreading of a liquid over a solid substrate, the contact line can stay pinned at sharp edges until the contact angle exceeds a critical value. At (or sufficiently near) equilibrium, this is known as Gibbs' criterion. Here, we show both experimentally and theoretically that for completely wetting volatile liquids there also exists a dynamically-produced critical angle for depinning, which increases with the evaporation rate. This suggests that one may introduce a simple modification of the Gibbs' criterion for (de)pinning, that accounts for the non-equilibrium effect of evaporation.
\end{abstract}
\maketitle

\section*{Introduction}

\begin{figure}[h!]
\includegraphics[scale=1.08]{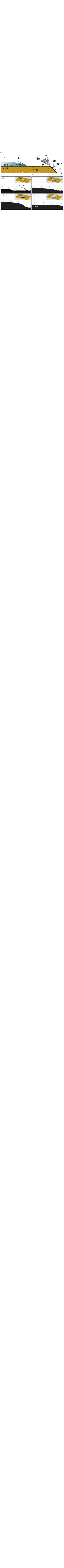}
\caption{Top: illustration of Gibbs' criterion for a droplet on a substrate with a sharp edge. The shaded zone indicates a range of possible equilibrium values (between $\theta$ and $\theta+\alpha$) that the contact angle can adopt when the contact line is pinned at the edge. Bottom: experimentally-obtained side views showing a part of a quasi-steadily evaporating droplet at different stages together with schematic representations (insets); (a) free, (b-c) pinned at the groove edge and (d) depinned contact line.}
\label{idea}
\end{figure}

Geometrical features on the surface of a rigid substrate, such as small-scale steps, grooves or other defects, pose an energy barrier hindering the motion of droplets or liquid films. The current understanding of this contact line pinning/depinning phenomenon is based on the equilibrium Gibbs' criterion \cite{gibbs}, which is graphically illustrated in Fig.~\ref{idea}-top in the case of a single sharp edge on an otherwise smooth substrate. In particular, depinning of a contact line from the edge occurs when its apparent contact angle $\theta_{ap}$ with respect to the horizontal exceeds the critical value 
\begin{equation}
\theta_{cr} = \theta + \alpha,
\label{gibbs}
\end{equation}
where $\alpha>0$ measures the downward slope of the solid past the edge, and $\theta$ is generally taken as the equilibrium Young's angle characterizing the given liquid/solid pair. 

In this Letter, we examine whether Eq.~\ref{gibbs} more generally holds out-of-equilibrium, i.e. when the contact angle $\theta$ is influenced (or even entirely determined) by hydrodynamics or heat transfer. Indeed, despite the fact that contact line pinning has been extensively studied for years \cite{mason77,bayramli,mason82,extrand,yeomans09,yeomans09_2,kalliadasis10,kalliadasis11,toth11,amirfazli}, to our knowledge such crucial question remains entirely open. Besides the impact this could have on our current understanding of contact angle hysteresis (for a review see \cite{quere08}), answering it could also result in major advances in several applications (e.g. microfluidics \cite{seemann08}, condensation on structured substrates \cite{enright}, dewetting \cite{ondar05}, nanowire growth \cite{tersoff, oh}, ...)  where heat, mass and momentum transfer processes could deeply affect pinning/depinning criteria. 

Here, we focus on the case of evaporating liquids, for which it is well known that even in perfect wetting conditions on atomically flat substrates, one can observe finite contact angles \cite{caza03,caza07,bonn09}. From the theoretical point of view, such evaporation-induced contact angles are now well understood on the basis of lubrication-type models \cite{morris01,pierre09,uwe12}, and known to occur even if the contact line is motionless (quasi-steady evaporation regime).  Indeed, a replenishing flow of liquid from the bulk is generally needed in this case, due to the intensification of evaporation near the contact line (where heat/mass transfer resistances are smaller). As this flow is driven towards a narrowing space, the system somehow tends to `mimimize' the overall viscous dissipation by increasing the apparent contact angle. In the present Letter, we demonstrate both experimentally and theoretically that such non-equilibrium angles (indeed increasing with the evaporation rate) may actually be used in place of the Young's angle in Eq.~\ref{gibbs}, hence providing a simple generalization of Gibbs' criterion for (de)pinning. Common intuition might suggest this should necessarily be so, but we here show that this actually depends on how localized the fluid flow is near the contact line, or in other words to what extent the apparent contact angle is independent of macroscopic length scales such as the droplet size, or the distance between the contact line and the edge.
\section*{Experimental Results}

\subsection*{A. Critical Angle Measurements}

\begin{figure}
\includegraphics[scale=1]{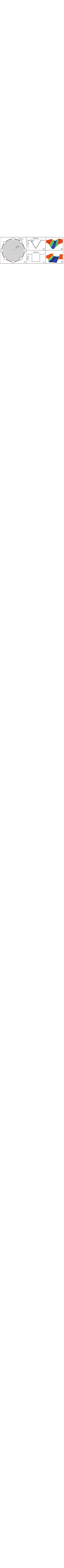}
\caption{(a) Several positions of the mask with respect to its circular path; (b) when the trajectory of the mask is aligned with its diagonal a V-shaped cross section is generated whereas (c) a U-shaped profile is observed when the trajectory is aligned to the sides of the mask; (d,e) 3D images of regions that correspond to triangular and rectangular profiles, respectively, taken with a laser confocal microscope.}
\label{groove}
\end{figure}

To this aim, we have conducted open-air experiments by continuously growing droplets of several volatile wetting liquids on a defined region of a 3 $mm$ thick polycarbonate slab (see Fig.~\ref{idea}-bottom). To work in a quasi-steady regime ruling out any effect of contact line speed, the capillary number was kept small ($Ca \leq 10^{-5}$) by working at sufficiently small injection rates. The region on which the droplets are grown is delimited by a circular microgroove 
of 5 $mm$ radius made on the solid substrate using an excimer laser. To determine the groove angle $\alpha$ we performed surface measurements with both an optical profilometer and a laser scanning confocal microscope. The acquired images indicate that the profile, and consequently the geometrical angle too, varies gradually along the groove owing to the excimer process. In particular, during the fabrication of the groove a square mask is placed between the substrate and the source of the excimer beam in order to define the area that will be exposed to the laser. Since the mask retains its orientation as it follows a circular path (Fig.~\ref{groove}a), cross sections of different shapes are produced (Fig.~\ref{groove}b, c). Yet, according to Eq.~\ref{gibbs} the depinning angle should occur at the cross-section characterized by the smallest slope. The results indicate that the minimal value of the geometrical angle is $\alpha=36.6$$\pm 1^o$. 

Placing the sample in a tensiometer (Kruss FM40 EasyDrop) enables one to follow the drop from its creation till depinning and to measure its apparent macroscopic contact angle $\theta_{ap}$ with respect to the horizontal. In the present study we focus on the critical angle $\theta_{cr}$ at which depinning occurs. A typical experiment is depicted in Fig.~\ref{idea}-bottom. In a first stage, liquid has just been deposited on the substrate (Fig.~\ref{idea}a), and with further liquid addition the droplet grows and gets flattened by gravity. At some moment, the contact line is pinned at the inner edge of the groove (Fig.~\ref{idea}b), and remains so while its contact angle increases until it reaches the critical angle for depinning (Fig.~\ref{idea}c). This moment is easily detected as the ``puddle" then rapidly floods the rest of the substrate (Fig.~\ref{idea}d). Note that only half of the drop is presented to achieve the largest possible magnification. For each experiment the sample is placed on the tensiometer with a different orientation to average over different parts of the groove. 

Three liquids of the same family (Hydrofluoroethers by 3M\texttrademark) differing mostly in their volatility are tested, conducting ten independent tensiometer experiments for each of them. In particular, we compare the depinning process of an extremely volatile liquid (HFE-7100) with a fairly volatile (HFE-7200) and a nearly non-volatile one (HFE-7500) (see Table \ref{table1}). Despite their very low surface tension, which points to perfect wettability on almost all kinds of surfaces, the results show that the microgroove can indeed pin the contact line, as expected \cite{bayramli}. However, the here found critical contact angles $\theta_{cr}^{exp}$ (Table~\ref{table1}) are not only larger than the geometrical angle $\alpha=36.6$$\pm 1^o$, but do also depend on the liquid (the difference $\theta_{cr}^{exp}-\alpha$ reaches about $8^o$ in the most volatile case).

\begin{table*}[top]
 \caption{\label{table1}Relevant properties of the tested liquids (equilibrium vapor pressure $P_{sat}$ and surface tension $\gamma$) along with experimentally-determined angles (critical angle at depinning $\theta_{cr}^{exp}$ measured by tensiometry, and interferometric measurements of the equilibrium contact angle $\theta_{eq}$ and the evaporation-induced angle $\theta_{ev}$). All values are obtained in ambient conditions.}
 \begin{tabular}{c c c c c c c} 
 Liquid & $P_{sat}$ & $\gamma$ & $\theta_{eq}$  & $\theta_{cr}^{exp}$ & $\theta_{cr}^{exp}-\alpha$ & $\theta_{ev}$ \\
HFE-7x00 & (kPa) &   (mN/m) &  ($^o$)  &  ($^o$) & ($^o$) & ($^o$)\\\hline 
 7100 & 26.7 & 13.6 & $\leq$0.5 & 44.8 $\pm$ 1.5 & 8.2 $\pm$ 2.5  & 8.8 $\pm$ 1.0\\
 7200 & 14.5 & 13.6 & - & 42.2 $\pm$ 2.5 & 5.6  $\pm$ 3.5 & 6.0 $\pm$ 1.0\\
 7500 & 2.1  & 16.2 & $\leq$0.2 & 38.4 $\pm$ 1.8 & 1.8  $\pm$ 2.8  & 3.4 $\pm$ 0.5\\\vspace{-4.8mm}
\end{tabular}
\end{table*}

\subsection*{B. Evaporation-Induced Contact Angle Measurements}

\begin{figure}[h!]
\includegraphics[scale=1.4]{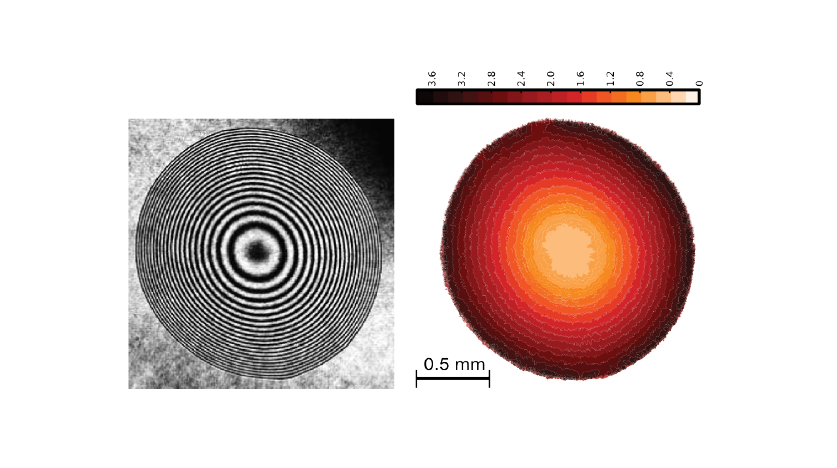}
\caption{Left: Typical fringe image of an evaporating sessile drop of HFE-7500 obtained by Mach-Zehnder interferometry. Right: Contour image of the extracted local angle of the drop after applying the 1D-CWT on the interference pattern.}
\label{idea2}
\end{figure}

\begin{figure}[br]
\includegraphics[scale=0.34]{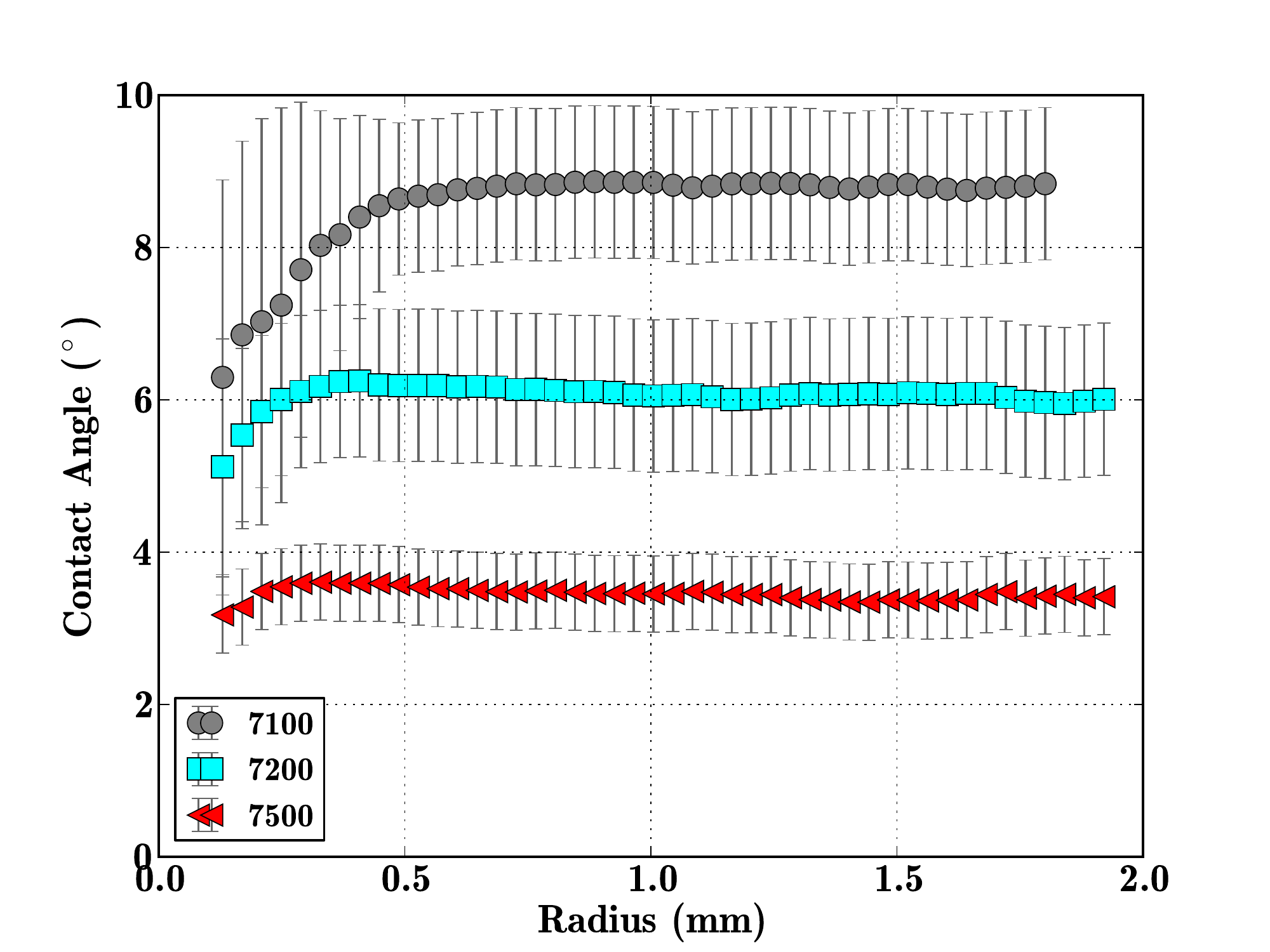}
\caption{Evaporation-induced contact angles versus base radius of the drop for the three different liquids studied (the error bars indicate the standard deviation of the contact angle along the drop periphery, with a 95\% confidence interval).}
\label{exp_plot}
\end{figure} 

To explore whether this difference is linked with evaporation, and to rule out any possibility of a non-vanishing Young's angle at equilibrium (i.e., under saturated conditions), we perform an independent series of tests. The widely used interferometric techniques are here optimized to be applicable to visualize evaporating droplets of millimetric sizes whose apparent contact angle extends far beyond the ones reported in the previous studies \cite{caza03}. In particular, with the current setup (Mach-Zehnder interferometer) we are able to measure a contact angle up to 18$^{\circ}$ while maintaining at the same time a field of view of approximately 5 x 5 $mm$ (assuming a 2K sensor with 5.5 $\mu m$ pixel size). Although this configuration is different (receding contact line) from the one described previously (advancing contact line), the corresponding evaporation-induced contact angles are here assumed not to differ significantly.

In these experiments, a drop of about 2$\mu$l is deposited on a flat transparent polycarbonate surface (same material as used in the previous experiments) and allowed to evaporate freely. The Mach-Zehnder interferometer itself consists in a laser beam (632.8 nm) crossing a beam splitter resulting into two coherent beams. One beam works as reference and therefore is directed straight to our recording system (Micro-Nikkor 105 $mm$ on a Prosilica GE1050). The other one first traverses the liquid phase in a direction orthogonal to the drop base before ending up in the camera. Thus, during the free receding motion of the droplet, we obtain interference fringes (Fig.~\ref{idea2}-left) that are processed using a dedicated 1D continuous wavelet transform technique (1D-CWT) \cite{Dehaeck13}. At each point of the image, deviations from the initial interference pattern are linked to drop height variations, $\Delta t$, through the equation $n \Delta t/\uplambda = \Delta\Phi/2\uppi$, where $\lambda$, $\Delta\Phi$ and $n$ are the wavelength, the phase difference between reference and object beams and the refractive index of the liquid, respectively. As previously, the experiments are repeated ten times for each liquid. Eventually, we extract the actual drop profile and calculate for all the three cases the apparent contact angle, $\theta_{ev}$, induced by evaporation (Fig.~\ref{idea2}-right). As the contact line recedes, the angle remains rather constant without pinning, and as expected it is larger for the more volatile cases (Fig.~\ref{exp_plot}). Moreover, having the complete drop in the field of view allows us to extract the angle along all its periphery. This indicates how axisymmetric the receding is (in Table~\ref{table1} an azimuthal average is given). Similar experiments are also performed under saturated conditions (closed box preventing evaporation), and as expected the drop then spreads until it indeed reaches a vanishing equilibrium contact angle (see $\theta_{eq}$ in Table \ref{table1}).

Most importantly, Table \ref{table1} shows that the excess $\theta_{cr}^{exp}-\alpha$ measured by tensiometry very satisfactorily compares with the evaporation-induced angle $\theta_{ev}$ independently measured by interferometry. This clearly suggests that one may introduce a simple modification of the Gibbs' critical angle (Eq.~\ref{gibbs}) for depinning, by merely replacing $\theta$ by the evaporation-induced angle $\theta_{ev}$, i.e.
\begin{equation}
\theta_{cr} = \theta_{ev} + \alpha 
\label{modgibbs}
\end{equation}

\section*{Theoretical Results and Discussion}

This experiment-based conjecture is now examined on the basis of a thin-film model valid for the case of a highly wetting liquid on a smooth superheated substrate. Given that the corresponding theory for diffusion-limited evaporation into air generally involves non-local operators \cite{eggers10}, we here consider the mathematically simplest problem of a droplet evaporating into its own vapor (\cite{morris01,pierre09,uwe12} and references therein). As our main purpose is to validate Eq.~\ref{modgibbs} on theoretical grounds, the actual value of $\theta_{ev}$ is not very important and the simplest model may indeed be used.

An essential feature of quasi-steady evaporation-induced angles is that they are generally determined within a so-called ``microstructure'' of the contact line, characterized by well-defined length scales, and largely independent of the particular macroscopic situation considered \cite{pierre09,pierre11}. To nondimensionalise the problem, it is thus convenient to use these microscales, that are based on a Hamaker constant $A$ quantifying the effective attraction of liquid molecules by the substrate, and on the superheat $\Delta T$ driving evaporation. Referring the reader to \cite{pierre09} for details, the vertical (i.e. film thickness) scale is defined by the thickness of the ultra-thin film in equilibrium with the vapor, i.e. $h_f=\left (A T_{\rm sat}/\rho {\cal L} \Delta T \right )^{1/3}$, where $T_{\rm sat}$ is the saturation temperature (at the given vapor pressure), $\rho$ is the liquid density, and $\cal L$ is the latent heat. Defining a molecular length scale by $a=\sqrt{A/\gamma}$, in which $\gamma$ is the surface tension, the horizontal length scale is given by $[x]=\epsilon^{-1} h_f$, where $\epsilon=\sqrt{3} a/h_f \ll 1$ is a parameter whose smallness underlies both the lubrication approximation ($h_f \ll [x]$) and the continuum assumption ($h_f \gg a$). Further, incorporating both a source term \cite{uwe12} that mimics a continuous liquid injection (as in our quasi-steady experiments) and a substrate topography $\eta(x)$ \cite{alex10, kalliadasis10}, the dimensionless steady-state equation for the film thickness $h(x)$ (i.e. the liquid/vapor interface is at absolute height $z=\eta+h$, see Fig.~\ref{idea}-left) is
\begin{equation}
-\frac{\partial q}{\partial x}-E \,j+S = 0
\label{eq_lubri}
\end{equation}
where the horizontal volumetric flux $q(x)$, the evaporation flux density $j(x)$, and the source $S(x)$ are given by
\begin{equation}
q=\frac{h^3}{3} \frac{\partial \Omega}{\partial x}~,~~~~j=\frac{1-\Omega}{K+h}~,~~~~S = f \exp \left [-\frac{x^2}{2 \sigma^2}\right ]
\label{def_qjS}
\end{equation}
and $\Omega$ is (minus) the excess pressure (including both disjoining and capillary components)
\begin{equation}
\Omega=\frac{1}{h^3}+3 \frac{\partial^2}{\partial x^2} (h+\eta)=h^{-3}+3 (h''+\kappa)
\label{Omega}
\end{equation}
where $\kappa(x)=\eta''(x)$ is the substrate curvature. In Eqs.~\ref{eq_lubri} and \ref{def_qjS}, 
$E$ and $K$ are the evaporation number and the dimensionless kinetic resistance, defined 
by $E=\mu\, \lambda \, T_{\rm sat}/3 (a \, {\cal L} \, \rho)^2$ and $K=\lambda \, T_{\rm sat}^2/L_{ww} \, {\cal L}^2 h_f$,
where symbols not yet defined are $\mu$ for the liquid dynamic viscosity, $\lambda$ for its thermal conductivity, and
$L_{ww}$ a phenomenological coefficient usually estimated by kinetic theory (see e.g. \cite{pierre11}). 

\begin{figure}[h]
\centering
\includegraphics[width=0.47\hsize]{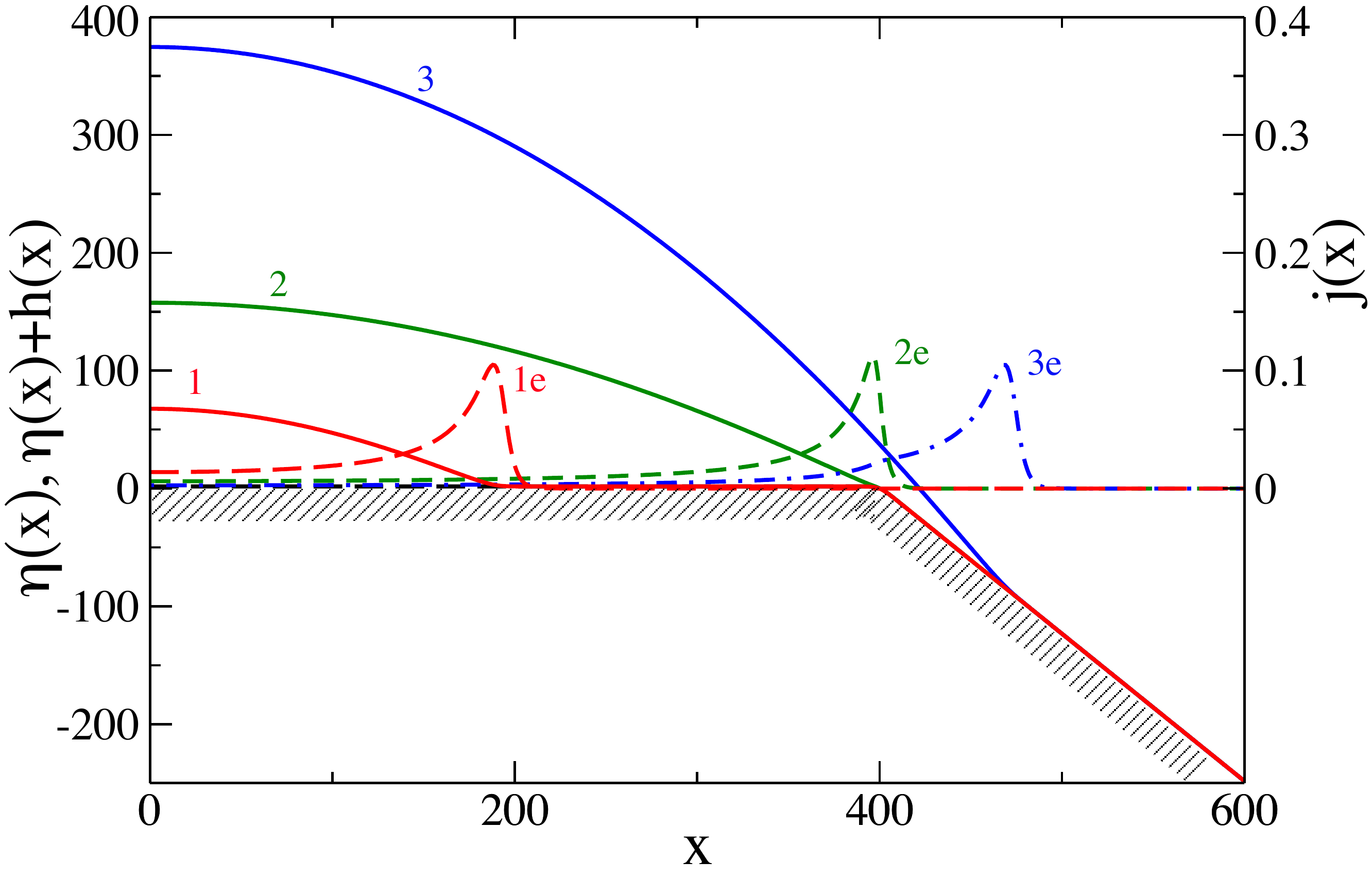}
\includegraphics[width=0.47\hsize]{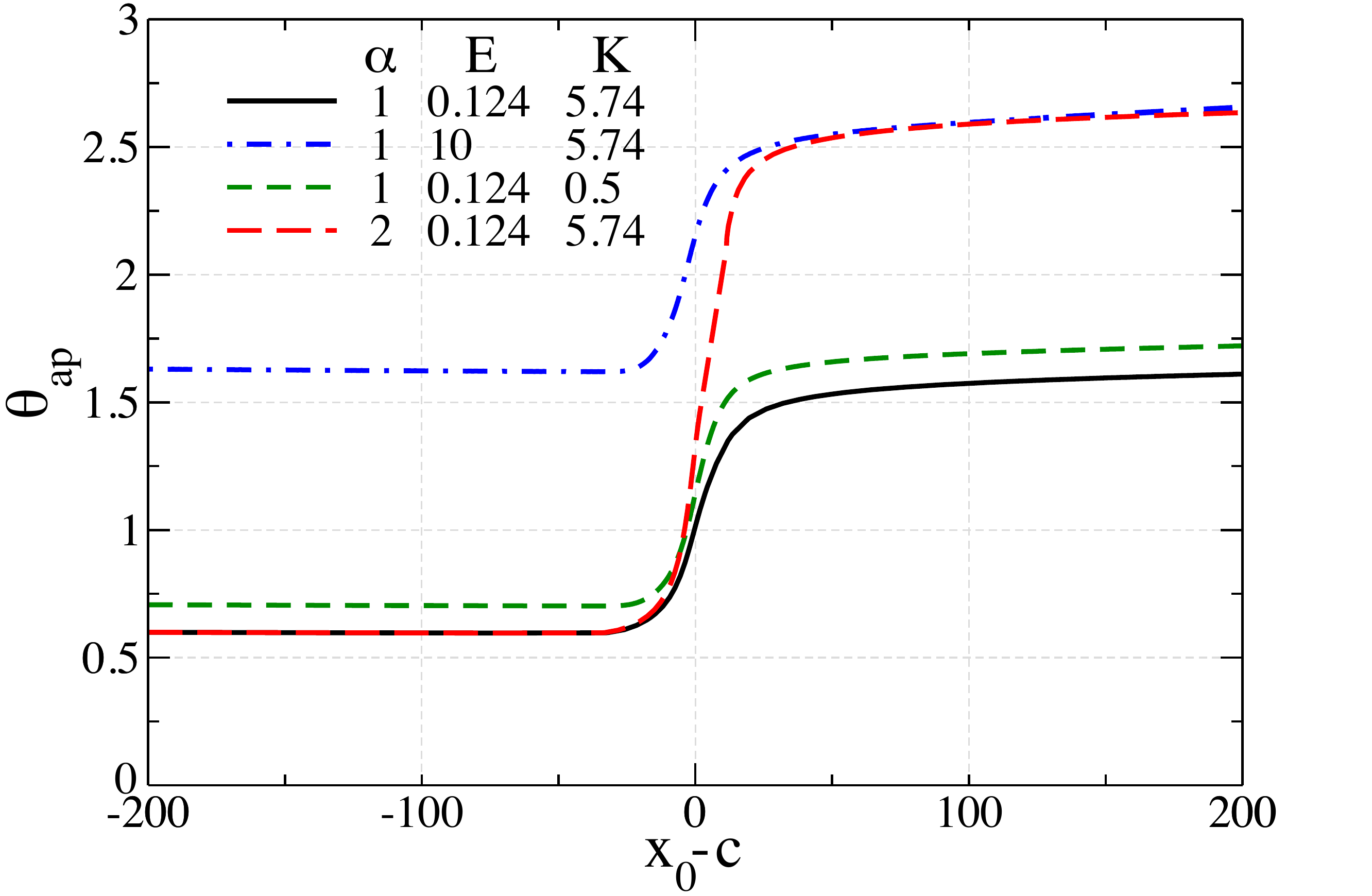}
\caption{Steady evaporating drops of a perfectly wetting
  liquid on a substrate with a downward bend. The left panel shows the
  bend profile $\eta(x)$ (downward angle $\alpha=1$) and typical droplet
  height profiles $\eta(x)+h(x)$ (solid lines; labeled 1, 2, 3; left
  scale) and the corresponding evaporation flux profiles
  $j(x)$ (broken lines; labeled 1e, 2e, 3e; right
  scale) for influx strength $f=0.075,0.08$ and $0.09$, respectively. Curve 1 is before pinning, curve 2 has a pinned
  contact line, and curve 3 is after depinning. Parameters are
  $E=0.124$, $K=5.74$, $w=5$, $\sigma=10$, $c=400$ and domain size $L=1000$.
The right panel gives the apparent angle $\theta_{ap}$ with respect to the
horizontal versus the position of the contact line $x_0$
(relative to the bend position $c$), for various cases as indicated in the legend.
The case $E=0.124$, $K=5.74$, $\alpha=1$ is the reference case shown in the left panel, 
for which $\theta_{ev} \simeq 0.57$ \cite{pierre09}. Remaining parameters are as in the left panel.}
\label{numresults}
\end{figure}

To model the inner edge of our groove, we consider a slope kink $\eta'(x)=-\alpha (1+\tanh{[(x-c)/w]})/2$, separating a horizontal region ($x \ll c$) from a region with downward slope $\alpha$ (for $x \gg c$). Hence, the curvature is given by $\kappa(x)=-\alpha\, {\rm sech}^2  [(x-c)/w]/2 w$, i.e. peaking in a region of width $\sim w$ around the bend at $x=c$. Boundary conditions are $h'(0)=h'''(0)=0$ (symmetry at $x=0$), while at $x=L \gg c$, we impose a flat equilibrium microfilm, i.e. $h(L)=1$ and $h'(L)=0$. This 4-th order problem is solved using the auto07p continuation package \cite{DoKK1991ijbc,DWCD2014ccp}, using typical values $E=0.124$ and $K=5.74$ as a reference case \cite{pierre09}. Selected typical results are presented in Fig.~\ref{numresults}.

The left plot shows that small droplets indeed display a non-vanishing evaporation-induced angle $\theta_{ev}$, despite the perfectly wetting situation \cite{uwe12}. The value of this apparent angle (defined as the slope at $z=0$ of a parabola fitting the droplet at its apex) is read at the right plot for sufficiently negative $x_0-c$ ($x_0$ is where the parabola intersects $z=0$). One notes that, $\theta_{ev}$ increases with $E$ and decreases with $K$, as parametrically studied in \cite{pierre09}. When increasing the influx strength $f$, the droplets increase in size and eventually reach the edge where their contact line remains pinned, while the apparent angle increases. At larger influxes, the droplet eventually depins and spreads on the sloped part of the substrate, with an apparent contact angle now close to $\theta_{ev}+\alpha$ (see right plot, now for positive $x_0-c$). This apparently holds whatever are the values of $E$, $K$, and $\alpha$, which  numerically validates our improved Gibbs' criterion (Eq.~\ref{modgibbs}). Note that our theory predicts a \emph{continuous} transition of the apparent contact angle between its extreme values $\theta_{ev}$ and $\theta_{ev}+\alpha$ when the contact line crosses the edge (see Fig.~5-right). This mesoscopic effect is also predicted by equilibrium density-functional theory \cite{dietrich}, due to the existence of an ultra-thin liquid film fully covering the substrate. In general, the fact that the corner is smooth rather than sharp (indeed $\eta(x)$ is a smooth function) also contributes to rendering the transition continuous. However, it appears here that the width of this ``pinning" region is rather determined by the size of the microregion in which the evaporation-induced contact angle is established,  i.e. typically the width of the evaporation flux peaks in Fig.~5-left. Note in this respect the slow (algebraic\cite{pierre09}) decay of the evaporation flux toward the drop center on Fig.~5-left, responsible for the fact that the apparent contact angle $\theta_{ap}$ tends only slowly to $\theta_{ev}+\alpha$ past the edge. Yet, the actual dimensional width of the evaporation flux peak is of the order of 35[x] $\approx$ 0.35$\mu$m for the reference case (E = 0.124, K = 5.74) considered here (see also the tabulated results in\cite{pierre09}). This therefore provides a characteristic length scale which both the droplet size and the distance of the contact line from the edge must significantly exceed for the transition to appear discontinuous, as in Gibbs' macroscopic picture.

\section*{Outlook}

To conclude, we have shown, both experimentally and theoretically, that evaporation enhances pinning of contact lines at sharp edges, even for the case of perfectly wetting liquids considered here. The critical angle to be reached for depinning is augmented above the value expected from equilibrium thermodynamics by a value increasing with the evaporation rate. To account for this effect, we have suggested to replace the Young's angle in the classical Gibbs' criterion by an angle dynamically induced by evaporation, and determined within some microscale vicinity of the contact line. Although the comparison between the experiments and the theory is not 
one to one, the latter has allowed to confirm this view, provided this microscale is sufficiently small compared to characteristic macroscales.

Further work should consider partially wetting situations (for which Young's angle can be increased by evaporation or decreased by condensation), and extend current theories to account for diffusion-limited evaporation into air. Besides, a widely open and very challenging perspective would be to study the impact phase change has on contact angle hysteresis of rough substrates.

\paragraph*{Acknowledgements:}

Financial support of FP7 Marie Curie MULTIFLOW Network (PITN-GA-2008-214919), ESA/BELSPO-PRODEX, BELSPO-$\mu$MAST (IAP 7/38) \& FRS-FNRS is gratefully acknowledged.

\bibliography{biblio}
\end{document}